\documentclass[a4paper]{jpconf}
\usepackage{psfrag}
\usepackage{graphicx}
\usepackage{amssymb}

\begin{document}
\title{Fermi level alignment in single molecule junctions and its dependence on interface structure}
\author{R. Stadler}
\address{Center for Atomic-scale Materials Physics, Department of Physics \\
  NanoDTU, Technical University of Denmark, DK-2800 Kgs. Lyngby, Denmark}

\date{\today}

\begin{abstract}
The alignment of the Fermi level of a metal electrode within the gap of the highest occupied and lowest unoccupied orbital of a molecule is a key quantity in molecular electronics. Depending on the type of molecule and the interface structure of the junction, it can vary the electron transparency of a gold/molecule/gold junction by at least one order of magnitude.
In this article we will discuss how Fermi level alignment is related to surface structure and bonding configuration on the basis of density functional theory calculations for bipyridine and biphenyl dithiolate between gold leads. We will also relate our findings to quantum-chemical concepts such as electronegativity.
 \end{abstract}

\begin{section}{Introduction}\label{sec:intro}
Interest for electron transport in nano-scale  contacts has recently intensified, because (i) the advent of the technologically motivated field of molecular electronics, where the use of organic molecules as the active parts in electronic devices is envisioned,~\cite{aviram74,joachim00} (ii) recent progress in the experimental techniques for manipulating and contacting individual molecules~\cite{reed97,smit02}, and (iii) the availability of first principles methods to describe the electrical properties of single molecule junctions on an atomic scale with high accuracy.~\cite{transiesta,ratner} These latter methods are usually based on density functional theory (DFT) in combination with a Keldysh formalism.

The alignment of the Fermi level of a metal electrode within the gap of the highest occupied (HOMO) and lowest unoccupied orbital (LUMO) of a molecule is a key quantity in molecular electronics, which can vary the electron transparency of a single molecule junction by an order of magnitude. In a recent publication \cite{first} we presented a quantitative analysis of the relation between this level alignment and charge transfer for bipyridine and biphenyl dithiolate (BPDT) molecules attached to gold leads based on density functional theory (DFT) calculations. The aim of the current work is to extend this study by analyzing the effect of varying surface structure and bonding configuration, where we also relate our findings to the quantum-chemical concept of electronegativity.
\end{section}

\begin{section}{Method}\label{sec:method}

All electronic structure calculations in this study are performed using a plane wave implementation of DFT~\cite{dacapo} with an energy cutoff of 340 eV, where the effect of the core on the valence electrons is approximated by ultra-soft pseudopotentials~\cite{vanderbilt90}, and exchange and correlation potentials and energies are parametrized as a PW91 functional.~\cite{pw91} 
In our calculations the supercells for the central region are defined by $3\times 3$ atoms in the directions perpendicular to the transport direction, and we used a $4\times 4$ $\bold k$-point grid in the transverse Brillouin zone. 

The conductance of the molecular junctions is calculated using a general non-equilibrium Green's function formalism for phase-coherent electron transport.~\cite{meir92}The Green's function of the central region, containing the molecule and three to four layers of the gold electrodes, is evaluated in terms of a basis consisting of maximally localized Wannier functions which are defined by an appropriate transformation of the Kohn-Sham eigenstates.~\cite{thygesen} The couplings to the semi-infinite gold electrodes are included via self-energy matrices which are also represented in the same Wannier function basis. 

\end{section}
  
\begin{section}{Transmission functions and MO level shifts}\label{sec:trans}

\begin{subsection}{Bipyridine}

  \begin{figure*}   
  \includegraphics[width=0.95\linewidth,angle=0]{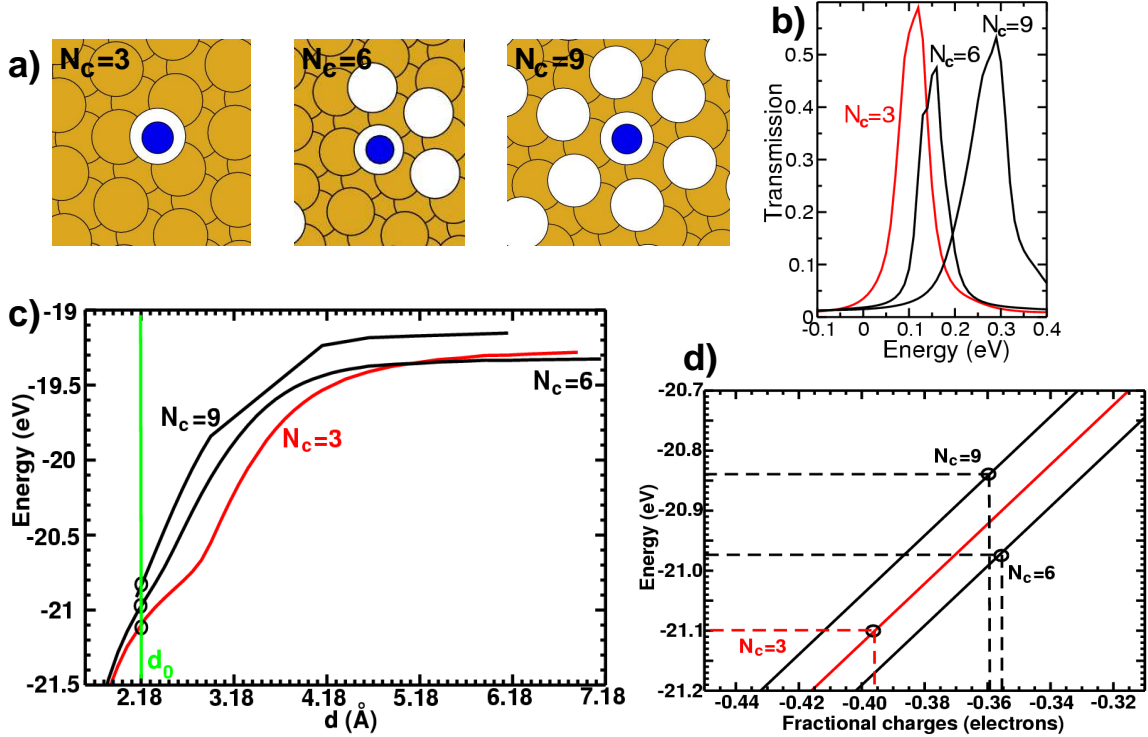}   
  \caption[cap.Pt6]{\label{fig.bipy_trans}(Color online) a) Variation of the surface contact geometry for a bipyridine molecule attached to a Au (111) surface, where $N_{c}$ (see text) varies from three to nine. Smaller dark spheres define the nitrogen positions. b) Transmission functions, c) energy shifts of MO1 (see text) depending on the distance $d$, where $d_0$ (2.18~\AA\ ) is the equlibrium bond length and d) charge dependent energetic positions of MO1 in an isolated molecule mimicking the interaction with the three different structures in a). All energies are given relative to the Fermi energy of the Au surfaces.      
}    
  \end{figure*}     

In a recent article~\cite{stadler} we presented a theoretical analysis of the conductance of a bipyridine molecular junction inspired by measurements on the same system.~\cite{xu} Since the atomic configuration of the electrode's surfaces were not known, we scanned a wide range of surface structures and found a strong dependence of our results on their geometries but overall agreement between theory and experiment in orders of magnitude ($\sim$ 0.01 $G_0$). All our calculations have been performed for an on-top contact of a bipyridine molecule on a Au (111) surface and as parameter for the variation of the surface geometry we used the coordination number $N_{c}$ (only including other Au atoms) for the surface atom directly connected to the molecule (see Fig.~\ref{fig.bipy_trans}a for three examples).
It was also found that the variation of the conductance of such bipyridine junctions with $N_{c}$ is entirely controlled by the position of the transmission peak due to the LUMO (Fig.~\ref{fig.bipy_trans}b) and that the effect of the interference with other orbitals as well as through-vacuum tunneling are negligible. 

But what is governing the LUMO position? We investigate the variation of the energetic position of the lowest-lying molecular orbital (denoted MO1 in the following text) with respect to the metal's Fermi level, in dependence on the distance between the surface and the molecule $d$ (Fig.~\ref{fig.bipy_trans}c).
Since MO1 is $\sim$ 10 eV below the lowest-lying Au valence states, its position must be exclusively guided by rigid potential shifts. Due to the quantitative correspondence of the sequence of the LUMO peaks in the transmission functions in Fig.~\ref{fig.bipy_trans}b to the energies of MO1 at $d_{0}$ (Fig.~\ref{fig.bipy_trans}c), hybridisation effects must be independent of $N_{c}$.
For large $d$, MO1 rests at an energetic position which exactly corresponds to the one it would hold due to vacuum level alignment (where the surface structures differ from each other depending on their respective work functions), which means that both subsystems are entirely charge neutral. 

Now we want to address the question how the rigid potential shifts that MO1 experiences at the bonding distance $d_{0}$ are related to charge being transferred between the molecule and the surface. For this purpose, in DFT the concept of fractional charges can be introduced.~\cite{parr} This makes it possible to determine the ground state electron density and electronic eigenvalue spectrum for a (albeit only finite) system with fractions of electrons removed or added when compared to the total charge of all the nucleis. Fig~\ref{fig.bipy_trans}d shows MO1 for isolated bipyridine cations in the region where 0.3 to 0.46 electrons have been removed from the molecule. The different lines for different values of $N_{c}$ result because the molecules vacuum level has to be aligned with the metals workfunction, which differs for the three cases. If Fig.~\ref{fig.bipy_trans}d is now compared with the MO1 positions at $d_{0}$, partial charges on the molecules can be attributed to different surface structures.

  \end{subsection}

  \begin{subsection}{Biphenyl dithiolate}
    \begin{figure*}
      \includegraphics[width=0.95\linewidth,angle=0]{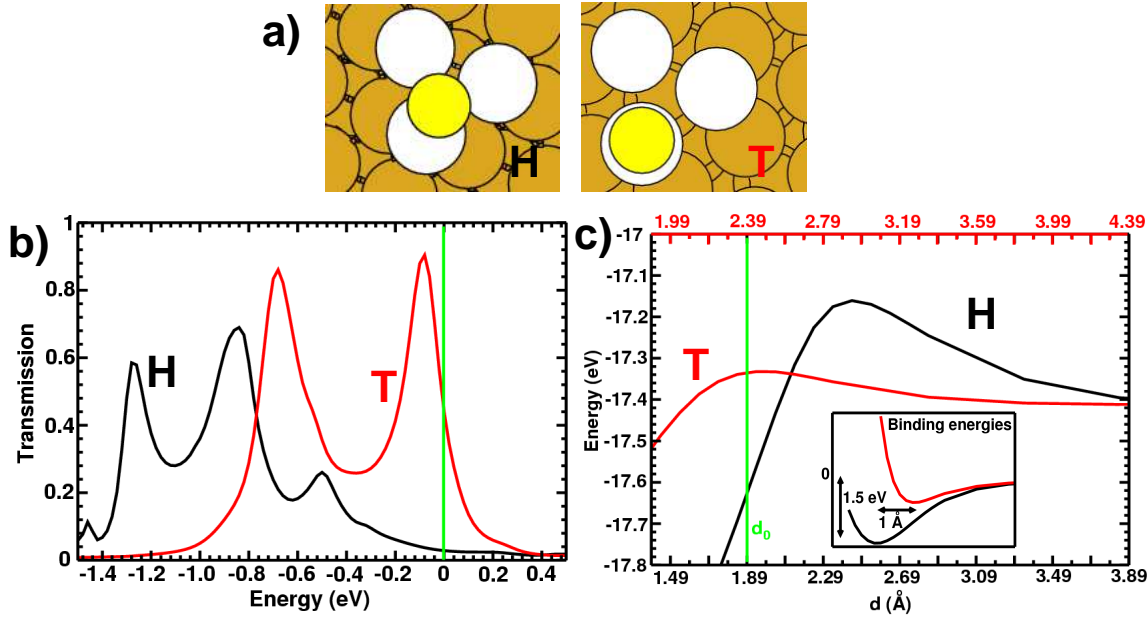}
      \caption[cap.Pt6]{\label{fig.BPDT_trans} (Color online) a) Contact geometry for a BPDT molecule connected to a triangle of Au atoms on a Au (111) surface. The small spheres denote the positions of the sulfur anchor atoms, where H stands for hollow and T for an on top configuration. b) Transmission functions and c) energy shifts of MO1 for the two configurations in a), where the distance axis is different for T and H in absolute values, since it is referring to the respective $d_0$ in both cases. The inset of c) gives the binding energies in dependence on the distance $d$ between the molecule and the surface. All energies are given relative to the Fermi energy of the Au surfaces.
      }
    \end{figure*}

For dithiolates there is an on going discussion whether the sulfur atoms of the molecule are attached to the Au substrate in a hollow (H) or on top (T) configuration (Fig.~\ref{fig.BPDT_trans}a). Although it is well known, that the hollow configuration is thermodynamically more stable, there are also arguments about kinetic effects during the formation of the junction, which would favour an on top geometry. 

In a comparison of the transmission functions (Fig.~\ref{fig.BPDT_trans}b) we find that although the double peak structure directly below the Fermi energy is qualitatively similar in both cases, its energetic distance from $E_F$ varies by $\sim$ 0.7 eV. As a direct consequence the conductance for T where $E_F$ lies at the shoulder of the peak, is 0.44 $G_0$, whereas for H crossing just the tails it goes down to 0.03 $G_0$. Inspecting the difference in MO1 shifts in Fig.~\ref{fig.BPDT_trans}c we find that at the bonding distance $d_0$, it is only $\sim$ 0.3 eV, which indicates that the influence of hybridisation on the position of the peaks in the transmission function is quantitatively different for the two geometries.
We also note that BPDT differs from bipyridine in that charge transfer occurs in the opposite direction. The amount of charge being moved at the bonding distance is by almost a factor of two smaller for BPDT, which is reflected in much smaller MO1 energy shifts in dependence on $d$. 

The most interesting feature in our study of BPDT is the difference in conductance between H and T configurations, which is caused by the energetic variation of the positions of the transmission functions in Fig.~\ref{fig.BPDT_trans}b and the related different lengths where the maxima occur in Fig.~\ref{fig.BPDT_trans}c with respect to $d_0$. When $d$ is reduced, the amount of charge transfer increases, which makes MO1 go up with respect to $E_F$. At the extremal point the molecule is so close to the surface that fractions of electrons are pushed out of the contact region, which is equivalent to a net loss of charge. For both, H and T, the position of the maximum of the MO1 vs. $d$ functions in Fig.~\ref{fig.BPDT_trans}c is at $d\sim$ 2.4. The two bonding configurations only differ in the sense that this distance also equals $d_0$ for T, whereas it is already in the repulsive regime for H (see for comparison also the distance dependent binding energies in the inset of Fig.~\ref{fig.BPDT_trans}c). Therefore it can be concluded that the large differences in conductance for the two geometries is an indirect consequence of their different bonding behaviours.

  \end{subsection}
\end{section}

\begin{section}{Electronegativity}

It is useful to look at our results from the perspective of a quantum chemical approach for calculating charge transfer, which is often used in the context of chemical reactions in organic or biochemistry~\cite{parr}. The key quantity in this scheme is the electronegativity $\mu$, and supplemented with the so called hardness $\nu$, it can be used for computing the total charge transfer between any two subsystems $A$ and $B$, when they are brought into contact, as 
\begin{equation}\label{eq:first}
\Delta N=(\mu_B - \mu_A)/2 (\nu_A + \nu_B).
\end{equation}
This concept has the appeal that no detailed information about single electron eigenenergies or hybridisation of orbitals needs to be analysed or even computed for the coupled system, and nevertheless quantitative estimates for the charge transfer can be made. Additionally it includes the relaxation of all electrons in the subsystems due to electron gain or loss as a consequence of calculating the ionisation potential and electron affinity from total energies of charged molecular fragments. 
As has been pointed out early~\cite{nalewajski}, however, the screening effects that act on the electrons on system $B$ due to the presence of $A$ and vice versa have been neglected in its derivation. As a consequence Eq.~\ref{eq:first} does not depend on the distance $d$ between the molecule and the surface, which is counterintuitive and in contradiction to our results. Recently, a semiempirical scheme has been used~\cite{tung} for a qualitative discussion of Fermi level pinning at metal-semiconductor interfaces and its dependence on the atomic structure of the interface. There a modified version of Eq.~\ref{eq:first}, which was corrected for its shape and distance dependency, has been derived,
\begin{equation}\label{eq:second}
\Delta N=\frac{\mu_B - \mu_A}{2(\nu_A + \nu_B)-2 J_{AB} + n_A J_{AA} +n_B J_{BB}}.
\end{equation}
In this expression $J_{AB}$ was a parameter for the electrostatic attraction between atoms from different sides of the interface, $J_{AA}$ and $J_{BB}$ denoted the repulsion between atoms within the same plane parallel to the interface and all three quantities were explicitly distance dependent. The number of next nearest neighbours within a plane was denoted $n_A$ and $n_B$ for the two subsystems, respectively. If we now compare Eq.~\ref{eq:second} with our results for the surface structure dependence of the charge transfer for the bipyridine system in Fig.~\ref{fig.bipy_trans}, we find that there is qualitative agreement in the sense that a small value of our coordination number $N_c$ corresponds to a large $n_A$ in Eq.~\ref{eq:second}, and both leads to an increased amount of transferred charge $\Delta N$.

  \end{section}

\begin{section}{Summary}\label{section:summary}

In summary, we discussed the level alignment in molecular nanojunctions for different Au surface structures in the case of bipyridine, and for different bonding configurations for BPDT.
We also compared our findings with the more traditional quantum chemical method of using electronegativities for characterising charge transfer, which in its simplest form is based on DFT total energy differences and not on single particle electronic eigenvalues. On a qualitative level the framework agrees well with our results for bipyridine in dependence on the Au surface structure.
\end{section}

\begin{section}{Acknowledgments}
I am very much indebted to Karsten W. Jacobsen for his encouragement and support throughout this project.
The Center for Atomic-scale Materials Physics is sponsored by the
Danish National Research Foundation. We acknowledge support from the
Nano-Science Center at the University of Copenhagen and from the
Danish Center for Scientific Computing through Grant No. HDW-1101-05.
\end{section}

\section*{References}
\bibliographystyle{apsrev}

\end{document}